\documentclass[superscriptaddress,showpacs,preprintnumbers,amsmath,amssymb,nofootinbib,twocolumn,prl]{revtex4}

\usepackage{graphicx}
\usepackage{dcolumn}
\usepackage{bm}
\usepackage{color}
\usepackage{wasysym}
\begin{document}

\title{Direct mass measurements of $^{19}$B, $^{22}$C, $^{29}$F, $^{31}$Ne, $^{34}$Na and other light exotic nuclei}

\author{L.~Gaudefroy}
\email{laurent.gaudefroy@cea.fr}
\affiliation{GANIL, CEA/DSM - CNRS/IN2P3, BP~55027, F-14076 Caen, France}
\affiliation{CEA, DAM, DIF, F-91297 Arpajon, France}
\author{W.~Mittig}
\affiliation{GANIL, CEA/DSM - CNRS/IN2P3, BP~55027, F-14076 Caen, France}
\affiliation{Department of Physics and Astronomy, Michigan State University, East Lansing, Michigan 48824, USA}
\author{N.A.~Orr}
\affiliation{LPC Caen, ENSICAEN, Universit\'e de Caen Basse-Normandie, CNRS/IN2P3, 14050 Caen, France}
\author{S.~Varet}
\affiliation{CEA, DAM, DIF, F-91297 Arpajon, France}
\author{M.~Chartier}
\affiliation{Oliver Lodge Laboratory, University of Liverpool, Liverpool L69 7ZE, United Kingdom}
\author{P.~Roussel-Chomaz}
\affiliation{GANIL, CEA/DSM - CNRS/IN2P3, BP~55027, F-14076 Caen, France}
\author{J.P.~Ebran}
\affiliation{CEA, DAM, DIF, F-91297 Arpajon, France}
\author{B.~Fern\'andez-Dom\'{i}nguez}
\affiliation{Oliver Lodge Laboratory, University of Liverpool, Liverpool L69 7ZE, United Kingdom}
\author{G.~Fr\'emont}
\affiliation{GANIL, CEA/DSM - CNRS/IN2P3, BP~55027, F-14076 Caen, France}
\author{P.~Gangnant}
\affiliation{GANIL, CEA/DSM - CNRS/IN2P3, BP~55027, F-14076 Caen, France}
\author{A.~Gillibert}
\affiliation{CEA-Saclay, DSM/IRFU/SPhN, F-91191, Gif-sur-Yvette Cedex, France}
\author{S.~Gr\'evy}
\affiliation{GANIL, CEA/DSM - CNRS/IN2P3, BP~55027, F-14076 Caen, France}
\author{J.F.~Libin}
\affiliation{GANIL, CEA/DSM - CNRS/IN2P3, BP~55027, F-14076 Caen, France}
\author{V.A.~Maslov}
\affiliation{Joint Institute of Nuclear Research, Dubna, Moscow region 141980, Russia}
\author{S.~Paschalis}
\affiliation{Oliver Lodge Laboratory, University of Liverpool, Liverpool L69 7ZE, United Kingdom}
\affiliation{Institut f\"ur Kernphysik, Technische Universität Darmstadt, 64289 Darmstadt, Germany}
\author{B.~Pietras}
\affiliation{Oliver Lodge Laboratory, University of Liverpool, Liverpool L69 7ZE, United Kingdom}
\author{Yu.-E.~Penionzhkevich}
\affiliation{Joint Institute of Nuclear Research, Dubna, Moscow region 141980, Russia}
\author{C.~Spitaels}
\affiliation{GANIL, CEA/DSM - CNRS/IN2P3, BP~55027, F-14076 Caen, France}
\author{A.C.C.~Villari}
\affiliation{GANIL, CEA/DSM - CNRS/IN2P3, BP~55027, F-14076 Caen, France}
\date{\today}

\begin{abstract}

We report on direct time-of-flight based mass measurements of 16 light neutron-rich nuclei.
These include the first determination of the masses of the Borromean drip-line nuclei
$^{19}$B, $^{22}$C and $^{29}$F as well as that of $^{34}$Na.  In addition, the most precise determinations 
to date for $^{23}$N and $^{31}$Ne are reported. Coupled with recent interaction cross-section measurements,
the present results support the occurrence of a two-neutron halo in $^{22}$C, with a dominant $\nu$2$s_{1/2}^2$
configuration, and a single-neutron halo in $^{31}$Ne with the valence neutron occupying predominantly the
2$p_{3/2}$ orbital. Despite a very low two-neutron separation energy the development of a halo in
$^{19}$B is hindered by the 1$d_{5/2}^2$ character of the valence neutrons.

\end{abstract}


\maketitle
Binding energies are of fundamental importance in nuclear physics and often provide, following 
the establishment of particle stability, the first means to explore the structure of a nucleus
and the nuclear landscape in general.
Indeed, one of the cornerstones of
nuclear structure, the magic numbers, was in part first recognized via irregularities in binding
energy systematics~\cite{MAYER}. More recently, ab initio studies of binding energies in 
the very lightest nuclei have uncovered the need to incorporate three-body forces in
the nuclear interaction~\cite{WIRINGA}.  
Here, we focus on the masses of light nuclei in the
vicinity of the neutron dripline, where recent theoretical advances have demonstrated
the key role of not only the three-body component in the nucleon-nucleon 
interaction~\cite{HAGEN,OTSUKA} but also the effects of the coupling to the 
continuum~\cite{OKOLOWICZ,TSUKIYAMA}.

Arguably the clearest case where the proximity of the continuum plays a role is 
that of the so-called halo nuclei, whereby the weak binding of valence nucleon(s) occupying
orbitals with low orbital angular momentum ($\ell \leq$~1) allows the development of
radial density distributions extending well beyond that of the ``core''~\cite{JENSEN}.
In the case of two-neutron halo nuclei, of which $^{11}$Li is the most well-known 
example and $^{14}$Be the heaviest firmly established case, the
systems are ``Borromean'': that is, whilst the full three-body system (core plus valence neutrons) 
is bound, neither of the two-body sub-systems are (core--neutron and neutron--neutron).

Changes in shell structure as the neutron dripline is approached are often key to halo
formation:  in the case of $^{11}$Li, for example, the valence neutron configuration exhibits a 
large 2$s_{1/2}^2$ component in addition to the naive expectation of 1$p_{1/2}^2$.
In terms of heavier halo systems, a recent reaction cross-section measurement of $^{22}$C,
the last bound $N$~=~16 nucleus, suggested the presence of a two-neutron halo with a valence 
neutron configuration dominated by $\nu$2$s_{1/2}^2$~\cite{TANAKA}, in line with the the existence of
shell closures at N=14 and 16~\cite{STANOIU2,OZAWA}, but in contradiction with the suggested reordering of the
$\nu$1$d_{5/2}$ and $\nu$2$s_{1/2}$ orbitals for $^{21}$C~\cite{STRONGMAN}.

Somewhat higher in mass, it has long be known that the $N=20$ shell closure
disappears for neutron-rich isotopes of Ne, Na and Mg~\cite{WARBURTON}.  This region, often referred to as the 
``island-of-inversion'', is characterised by ground-states dominated by $fp$-shell intruder configurations rather 
than the normal $sd$-shell configurations.  Very weakly bound nuclei lying within the island-of-inversion
with significant 2$p_{3/2}$ valence neutron configurations could, therefore, generate halos.
Indeed, recent measurements of the 
Coulomb dissociation and total reaction cross sections for $^{31}$Ne have lead to 
suggestions that it exhibits a single-neutron halo based on
a 2$p_{3/2}$ valence neutron~\cite{NAKAMURA,TAKECHI}. 
However, the very large uncertainty~\cite{JURADO} in its binding energy has precluded definitive
conclusions from being drawn.

In this Letter we report on direct time-of-flight (TOF) mass measurements of 16 
light neutron-rich nuclei lying at or close to the drip-line.  
These include the first experimental determination
of the masses of the Borromean systems $^{19}$B, $^{22}$C and
$^{29}$F as well as that for $^{34}$Na, and greatly improved masses for $^{23}$N and $^{31}$Ne.
The formation of halos in $^{19}$B, $^{22}$C and $^{31}$Ne is discussed in the light of these results.
  
The principle of the direct TOF technique is based on the relationship between the magnetic rigidity and
velocity, $B\rho=\gamma m_0 v/Q$, where $\gamma$ is the Lorentz factor, $Q$ is the atomic charge state,
$v$ the velocity and $m_0$ the rest mass. As such, a precise determination of the magnetic rigidity and 
velocity will enable the mass to be deduced~\cite{GILLIBERT86,GILLIBERT87}. 
In the present study, performed at GANIL (Grand
Acc\'el\'erateur National d'Ions Lourds, Caen, France), neutron-rich nuclei were produced via
fragmentation of a 4~$e\mu$A $^{48}$Ca$^{19+}$ beam of 60~A$\cdot$MeV on a rotating Ta
target mounted on a C backing. The Ta foil had slots machined in it such that for 
80\% of the beam exposure, the thickness was optimum for the
production of neutron-rich nuclei in the region of $^{22}$C. The
remaining 20\% provided for the production of less exotic nuclei with well-known 
``reference masses''.  The target was located between the solenoids of 
the SISSI device~\cite{ANNE} and the reaction products were selected in 
rigidity using the $\alpha$-shaped beam-analysis spectrometer and transported some 82~m to the focal 
plane of the high resolution
energy-loss spectrometer SPEG~\cite{BIANCHI} (Spectrom\`etre \`a Perte d'Energie du Ganil). The TOF of the secondary beam ions was measured
on a particle-by-particle basis between a thin-foil micro-channel plate detector located at the 
exit of the $\alpha$ spectrometer and a
plastic scintillator at the focal plane of SPEG. A drift
chamber located at the dispersive plane of SPEG was used to provide for a precise measurement of the
magnetic rigidity of each ion.
Nuclei transmitted to the SPEG focal plane were identified in $A$, $Z$ and $Q$ using standard
$\Delta$E--E--TOF techniques. The energy loss ($\Delta$E) was measured using an ionization
chamber and the total energy (E) was derived from a 1~cm thick 
BC400 scintillator. All the nuclei were 
determined to be fully stripped ($Z=Q$). The detection system did not allow any isomeric states
to be identified.  As such, those reference mass nuclei with known long lived isomeric
states were excluded from the analysis.

Two different settings of the alpha-spectrometer, beam line and SPEG were employed in the experiment. 
The first, with
$B\rho$~=~2.88~Tm, was optimized for mass measurements in the region of $^{22}$C. In a second shorter run,
with $B\rho$~=~2.4~Tm, heavier nuclei in the vicinity of $^{48}$Ca were studied with the aim of remeasuring the
mass of $^{47}$Ar for which contradictory results exist~\cite{AUDI,BENENSON,GAUDEFROY}.
Owing to the variable thickness of the production target, a very broad range of nuclei was 
populated for each setting, including, critically, reference mass nuclei for which at least 
three independent and coherent mass measurements have been reported.

The experimental masses were obtained using a fitting function $M(A,Z,m_0^\prime)$
of the form, 
\begin{eqnarray*}
M(A,Z,m_0^\prime) & = & m_0^\prime + C_1 + C_2 A + C_3 Z \\
& + & C_4 A^2 + C_5 Z^2 + f(\Delta E),
\end{eqnarray*}
where second order terms in $A$ and $Z$ as well as the function $f(\Delta E)$ are introduced 
to account for higher order corrections. The constants, $C_i$, were adjusted through a $\chi ^2$ 
minimization procedure to provide for the best possible reproduction of the reference masses. The 
variable $m_0^\prime$ is a first approximation for the mass derived directly from the measured TOF and $B\rho$. The new mass excesses 
deduced applying this procedure 
are reported in Table~\ref{TABLE1}. 

\begin{table}[t]
  \caption{\label{TABLE1}Measured mass excesses and uncertainties, in MeV,
    compared to results from previous work, including the 2003 Atomic Mass Evaluations~\cite{AUDI}.}
  \begin{ruledtabular}
    \begin{tabular}{cccc}
      Nucleus   & This work    & Ref~\cite{JURADO} & Ref~\cite{AUDI} \\
      \hline
      $^{19}$B  & 59.77~(0.35) &               &   \\
      $^{20}$C  & 37.36~(0.27) &               &  37.56~(0.24)  \\
      $^{22}$C  & 53.64~(0.38) &               &   \\
      $^{22}$N  & 31.11~(0.26) &               &  32.04~(0.19) \\
      $^{23}$N  & 36.72~(0.28) &  36.68~(0.86) &               \\
      $^{27}$F  & 25.45~(0.26) &  24.63~(0.19) &  24.93~(0.38) \\
      $^{29}$F  & 40.15~(0.35) &               &               \\
      $^{30}$Ne & 23.31~(0.28) &  23.04~(0.28) &  23.10~(0.57) \\
      $^{31}$Ne & 31.44~(0.31) &  30.82~(1.62) &               \\
      $^{33}$Na & 23.78~(0.30) &  23.42~(0.35) &  24.89~(0.87) \\
      $^{34}$Na & 31.68~(0.40) &               &   \\
      $^{43}$S  &-12.13~(0.22) & -12.07~(0.10) &  -11.97~(0.20)\\
      $^{44}$S  & \phantom{1}-9.34~(0.28) &  \phantom{1}-9.10~(0.14) & \phantom{1}-9.12~(0.39)\\
      $^{45}$Cl &-17.79~(0.22) & -18.36~(0.10) &  -18.36~(0.12) \\
      $^{46}$Cl &-14.01~(0.28) & -13.81~(0.16) &  -14.71~(0.72) \\
      $^{47}$Ar &-25.06~(0.19) &               &  -25.91~(0.10) \\

    \end{tabular}
  \end{ruledtabular}
\end{table}

The uncertainties were derived from the quadratic sum of three contributions~\cite{ORR}:
the statistical error;
{the systematic error -- 240~keV and 150~keV for the first and second settings, respectively -- and
the error estimated to result from extrapolating the
fitting function beyond the reference masses. The somewhat higher systematic error
for the first setting arose from the high count rates
associated with very light particles inducing pile-up in the SPEG focal plane detectors.

As may be seen in Table~\ref{TABLE1}, the masses of $^{19}$B, $^{22}$C, $^{29}$F,
and $^{34}$Na have been measured for the first time.
In the cases of $^{23}$N and $^{31}$Ne the precisions have been very substantially improved.
The present 
results are nearly all in good agreement with the 2003 mass evaluations~\cite{AUDI} and the
subsequent work of Jurado~\cite{JURADO}. 
In the case of $^{22}$N, the present result agrees with the first measurement by 
Gillibert~\cite{GILLIBERT87}, which was excluded from the mass evaluations~\cite{AUDI}, the listed 
value of which 
is based on the determinations of Refs.~\cite{ORR,WOUTERS}.  The present result for $^{45}$Cl is some
2.5$\sigma$ less bound than both Jurado {\em et al.} and the mass evaluations value (which was derived principally from the
work of Ref.~\cite{SARAZIN}).  In principle, such a result could arise from the population
of an isomeric state in $^{45}$Cl, however, Refs.~\cite{JURADO,SARAZIN} employed the fragmentation of 
$^{48}$Ca and a very similar setup to that used here.

The present mass excess for $^{47}$Ar deviates by more than 4$\sigma$ from the
value tabulated in the mass evaluations~\cite{AUDI}, which was derived directly from
Ref.~\cite{BENENSON} -- a heavy-ion multi-nucleon transfer reaction study -- where the ground state may
well have been masked by backgrounds arising from reactions other than that of interest.
The present mass excess is, however, in good agreement with that of -25.21(0.09)~MeV deduced
from the Q-value of the d($^{46}$Ar,p)$^{47}$Ar$_{gs}$ reaction~\cite{GAUDEFROY}.

\begin{figure}[!t]
\includegraphics[width=7.cm]{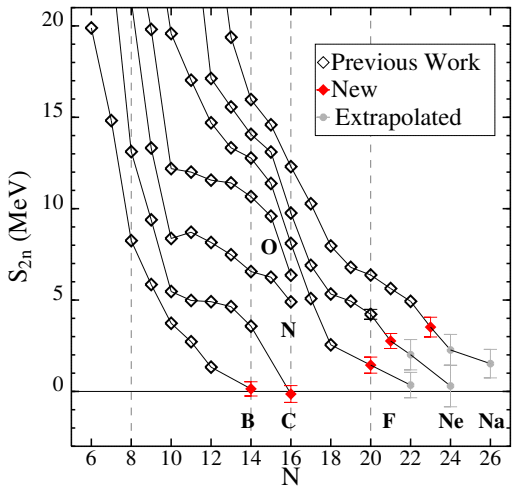}
\caption{\label{FIGURE1}(Color on line.)
  Two-neutron separation energies ($S_{2n}$) as a function of neutron number. Red symbols correspond to masses
  measured here for the first time and gray points to masses estimated on the basis of extrapolations~\cite{AUDI2011}. 
  The horizontal line delineates the limit of binding ($S_{2n}=0$~MeV), whilst vertical lines indicate known (sub-) shell closures.
  The lines connecting the points for each Z are to guide the eye.
}
\end{figure}

The two-neutron separation energies ($S_{2n}$) corresponding to the mass excesses tabulated in Table~\ref{TABLE1}
and including recently estimated extrapolations for unmeasured masses~\cite{AUDI2011}
are reported in Fig.~\ref{FIGURE1}.
It is clearly apparent that $^{19}$B and $^{22}$C are at the limit of binding -- $S_{2n}$= 0.14(39) and -0.14(46) MeV respectively --
thus fulfilling an essential condition for the development of a two-neutron halo.
Similarly, the very weak one-neutron binding energy of $^{31}$Ne and $^{34}$Na -- $S_n$=-0.06~(41)~MeV and
0.17~(50)~MeV, respectively -- suggest that single-neutron halos could develop. 
In the following the impact of the present measurements on the possible halo structure of
$^{19}$B, $^{22}$C and $^{31}$Ne is discussed (in the case of $^{34}$Na the necessary complementary reaction
cross section has yet to be measured).
 
We employ as a starting point a simple model, first proposed by Hansen and Jonson~\cite{HANSEN}, whereby the matter radius, $R_m$, may be estimated as follows,

\begin{equation*}
R_m^2(A) = \frac{A-k}{A} \big( R_c^2 + \frac{k}{A}r_\nu^2 \big), 
\end{equation*}

where $A$ is the mass number of the nucleus, $k$ the number of valence neutron(s), 
$R_c$ the matter radius of the core nucleus,
and $r_\nu$ the radius of the valence neutron distribution. 
The values of $R_c$ used here were $2.99(9)$~fm for $^{17}$B~\cite{SUZUKI}, $2.98(5)$~fm
for $^{20}$C~\cite{OZAWA2} and $3.14(11)$~fm for $^{30}$Ne. The latter value was deduced from the results 
of a recent interaction cross-section measurement~\cite{TAKECHI} as discussed below for the case of $^{31}$Ne.
The valence neutron radius, $r_\nu$, was estimated using a Woods-Saxon well ($a$~=~0.65~fm and $r_0$~=~1.25~fm), the 
depth
of which was varied in order to ensure the correct binding of the valence neutron with quantum
numbers $n$ and $\ell$. For the two-neutron halos, it was assumed that the binding energy is
shared equally between them.

\begin{figure}[!t]
\includegraphics[width=8.5cm]{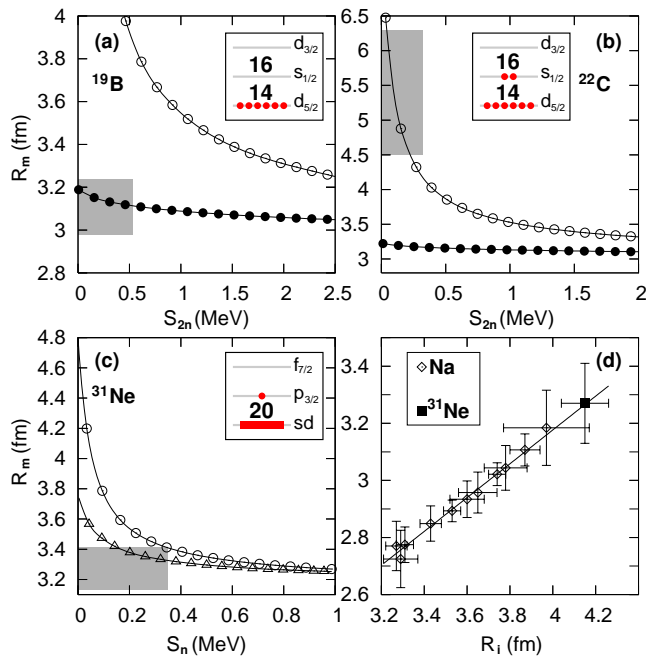}
\caption{\label{FIGURE2} Matter radii, $R_m$, for (a) $^{19}$B, (b) $^{22}$C and (c) $^{31}$Ne as a function of neutron separation
energy [$S_{(2)n}$]. Curves marked with \CIRCLE, \Circle $~$ and $\triangle$ correspond to calculations assuming $1d$, $2s$ and
$2p$ valence neutron configurations, respectively. The experimentally deduced $R_m$ and present mass determinations
are shown (shaded rectangles). The insets indicate the principal valence neutron configuration deduced for each nucleus.
Panel (d): $R_m$ for $^{22-31}$Na as a function of interaction radius ($R_I$) where the solid line represents a linear adjustment.
The filled square corresponds to the matter radius for $^{31}$Ne deduced from its interaction radius (see text).}
\end{figure}

The results of calculations over a range of valence neutron(s) binding energies are displayed, for various assumptions
for pure configurations, in Figs.~\ref{FIGURE2}~(a)-(c) along with the experimentally determined
$R_m$ and $S_{(2)n}$, where the matter radii for
$^{19}$B and $^{22}$C were taken from Refs.~\cite{SUZUKI} and ~\cite{TANAKA}, respectively.
In case of configuration mixing, the matter radius is simply the weighted mean of the radii for each configuration.
As may be seen in Fig.~\ref{FIGURE2}~(a) the results for $^{19}$B are well described when the valence neutrons are predominantly of a 
$d$-wave character.
Owing to the very extended nature of the $2s$ orbital, any admixture of $(2s_{1/2})^2$ configuration is expected
to be less than 20\%.
Consequently, the radial extent of the tail in the $^{19}$B valence neutron distribution will be strongly suppressed by the $d$-wave
angular momentum barrier.\\
On the contrary, the results for $^{22}$C strongly support a two-neutron halo with a predominantly $s$-wave
configuration, with the $d_{5/2}$ orbit being fully occupied [Fig.~\ref{FIGURE2}~(b)] and any appreciable contribution
from higher lying $d_{3/2}$ orbital being very unlikely. These conclusions are in line with that of Refs.~\cite{TANAKA,KOBAYASHI,FORTUNE}.
The results for $^{19}$B ($N$~=~14) and $^{22}$C ($N$~=~16) imply that the single-particle level ordering at the 
dripline for $N$~=~14 and 16 is similar to that near stability -- that is, the $\nu$1$d_{5/2}$ orbit lies below the $\nu$2$s_{1/2}$ orbit.
As such, the scenario of degeneracy or inversion of these two levels in $^{21}$C, as suggested based on the
observation of a possible 3$^-$ resonance in $^{22}$N~\cite{STRONGMAN}, is rather unlikely.

Although the interaction cross-section ($\sigma_I$) has been reported for $^{31}$Ne on a carbon 
target~\cite{TAKECHI}, its matter radius has
yet to be deduced.  Here we estimate
it based on $\sigma_I$($^{31}$Ne) and the systematics of the matter radius versus interaction radius, $R_I$, for the neighbouring isotopes $^{22-31}$Na derived by Ref.~\cite{SUZUKI2}.
An interaction radius of 4.15~(11)~fm was estimated for $^{31}$Ne using the relation~\cite{TANIHATA}, 
$\sigma_I=\pi [R_I(^{31}Ne)+R_I(C)]^2$, where $R_I(C)$~=~2.61~fm is the interaction radius of $^{12}$C and
$\sigma_I$($^{31}$Ne)~=~1435~(22)~mb~\cite{TAKECHI}.  As shown in Fig.~\ref{FIGURE2}~(d), by extrapolating a linear adjustment to the matter versus interaction radius for the Na isotopes, $R_m$($^{31}$Ne)~=~3.27~(14)~fm is deduced.
This value, in combination with the much improved mass for $^{31}$Ne, suggests [Fig.~\ref{FIGURE2}~(c)] that the valence neutron is predominantly 
of a $p$-wave character, a result reinforcing those of Refs.~\cite{NAKAMURA,TAKECHI,HORIUCHI,MINOMO} constrained by the earlier very imprecise mass.
We note that mixing between the configurations considered here is not possible owing to their different parities.
As discussed in Refs.~\cite{MINOMO,HAMAMOTO,URATA}, the inversion of the $\nu f_{7/2}$--$\nu p_{3/2}$ configurations in $^{31}$Ne is driven by deformation effects.

Whilst the mass of $^{31}$F was beyond the reach of the present work, an estimate 
derived from extrapolations~\cite{AUDI2011} suggests that
it is just bound, in line with its particle-bound character.  Coupled with
the mass obtained here for $^{29}$F, no downward trend in the $S_{2n}$ systematics indicative of a shell closure
at N=20 is apparent for Z=9 (Fig.~\ref{FIGURE1}).  This suggests that the island-of-inversion could extend to the lightest bound N=20
nucleus, $^{29}$F, as suggested theoretically~\cite{UTSUNO}.

In summary, new direct time-of-flight mass measurements of 16 light neutron-rich nuclei are reported.
Of these, the masses for $^{34}$Na and the Borromean systems, $^{19}$B, $^{22}$C and $^{29}$F, have been obtained for
the first time. In the cases of $^{23}$N and $^{31}$Ne, the masses have been determined with greatly improved
precisions. By combining these results with matter radii derived from interaction cross-section measurements 
the possible halo structures of $^{19}$B, $^{22}$C and $^{31}$Ne have been constrained.
In the case of $^{22}$C a two-neutron halo with a predominantly $\nu$2$s_{1/2}^2$ configuration is favoured, whilst,
despite a very low two-neutron binding energy, the development of a halo in
$^{19}$B is hindered by the 1$d_{5/2}^2$ character of the valence neutrons.  As such, the ordering at the 
neutron dripline of the
single-particle levels at N=14 and 16 follows normal shell model expectations.
The present, much more precise mass for $^{31}$Ne, supports the development of 
a single-neutron halo with the valence neutron occupying mainly the
2$p_{3/2}$ orbital. 
If deformation effects also result in
the valence neutron of $^{34}$Na ($S_n$~=~0.17$\pm$0.50~MeV)
being predominantly $2p_{3/2}$, then a measurement of the interaction cross section for the 
most neutron-rich Na isotopes would be of interest in establishing its possible halo character.

The authors wish to acknowledge the efforts of the GANIL operations group in maintaining SISSI
and for providing the stable high intensity $^{48}$Ca beam which was essential for the success of the experiment.

\end{document}